\documentclass[12pt]{iopart}
\usepackage{graphicx}
\begin{document}
\title{Random wave functions and percolation}

\author{E  Bogomolny and \fbox{ C  Schmit}}

\address{CNRS, Universit\'e Paris-Sud,\\ Laboratoire de Physique Th\'eorique et Mod\`eles Statistiques,\\ 91405 Orsay Cedex, France}

\ead{eugene.bogomolny@lptms.u-psud.fr}

\begin{abstract}
Recently it was conjectured that nodal domains of random wave functions are adequately described by critical percolation theory. In this paper we strengthen this conjecture in two respects. First, we show that, though wave function correlations decay slowly, a careful use of Harris' criterion confirms that these correlations are unessential and nodal domains of random wave functions belong to the same universality class as non-correlated critical percolation. Second, we argue that level domains of random wave functions are described by the non-critical percolation model.   
\end{abstract}
\maketitle

\section{Introduction}

In 1977 Berry \cite{berry} conjectured  that wave functions of generic chaotic systems, $\Psi(\vec{x}\,)$,  can statistically be written as random superposition of  a complete  set of functions, $\psi_m(\vec{x}\,)$,
\begin{equation}
\Psi(\vec{x}\,)=\sum_{m}C_{m}\psi_m(\vec{x}\,)
\label{berry}
\end{equation}
where the coefficients $C_m$ are independent identically distributed random variables with zero mean and variance obtained from normalization. 
In particular, any wave function  of a two-dimensional billiard obeys the Helmholtz equation with energy $E=k^2$
\begin{equation}
(\Delta +E)\Psi(x,y)=0
\end{equation}
and can be represented as a formal sum
\begin{equation}
\Psi(x,y)=\sum_{m} C_m J_{|m|}(kr){\rm e}^{{\rm  i} m\phi}\ ,
\label{bessel}
\end{equation}
where $r$ and $\phi$ are  polar coordinates  and  $J_m(r)$ are the usual Bessel functions. For problems without magnetic field wave functions  are real and   
$C_m=C_{-m}^{*}$. 

Consider a chaotic quantum system like the stadium billiard with  Dirichlet boundary conditions and let us  calculate (numerically) a large number of  eigenfunctions with energies close to $k^2$. Each eigenfunction  gives a well defined set of coefficients, $C_m$ in the expansion  (\ref{bessel}) and it is of interest  to know  mean values of different functions of these coefficients over the whole ensemble of eigenfunctions.    
Berry's conjecture means  that  in the semiclassical limit $k\to \infty$ the result  will be the same as if the coefficients $C_m$ would be  independent Gaussian random variables with 
\begin{equation}
\left <C_m\right >=0\;\;\;\mbox{ and }\;\left <C_mC_n^*\right >=\sigma^2\delta_{mn}\;.
\end{equation}
\begin{figure}[!b]
      \centering
      \includegraphics[width=.39\textwidth, angle=-90]{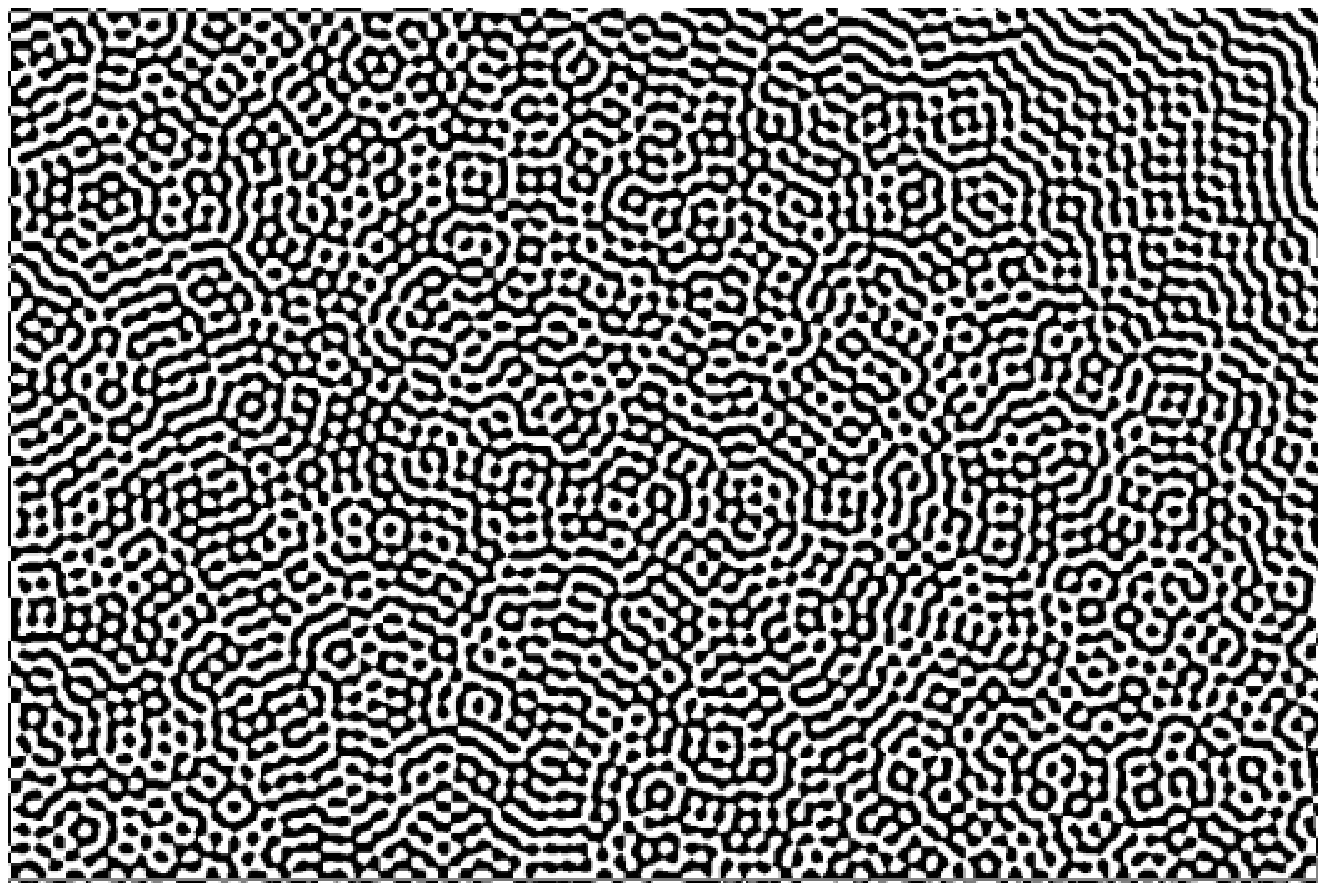}\hfill
      \includegraphics[width=.39\textwidth, angle=-90 ]{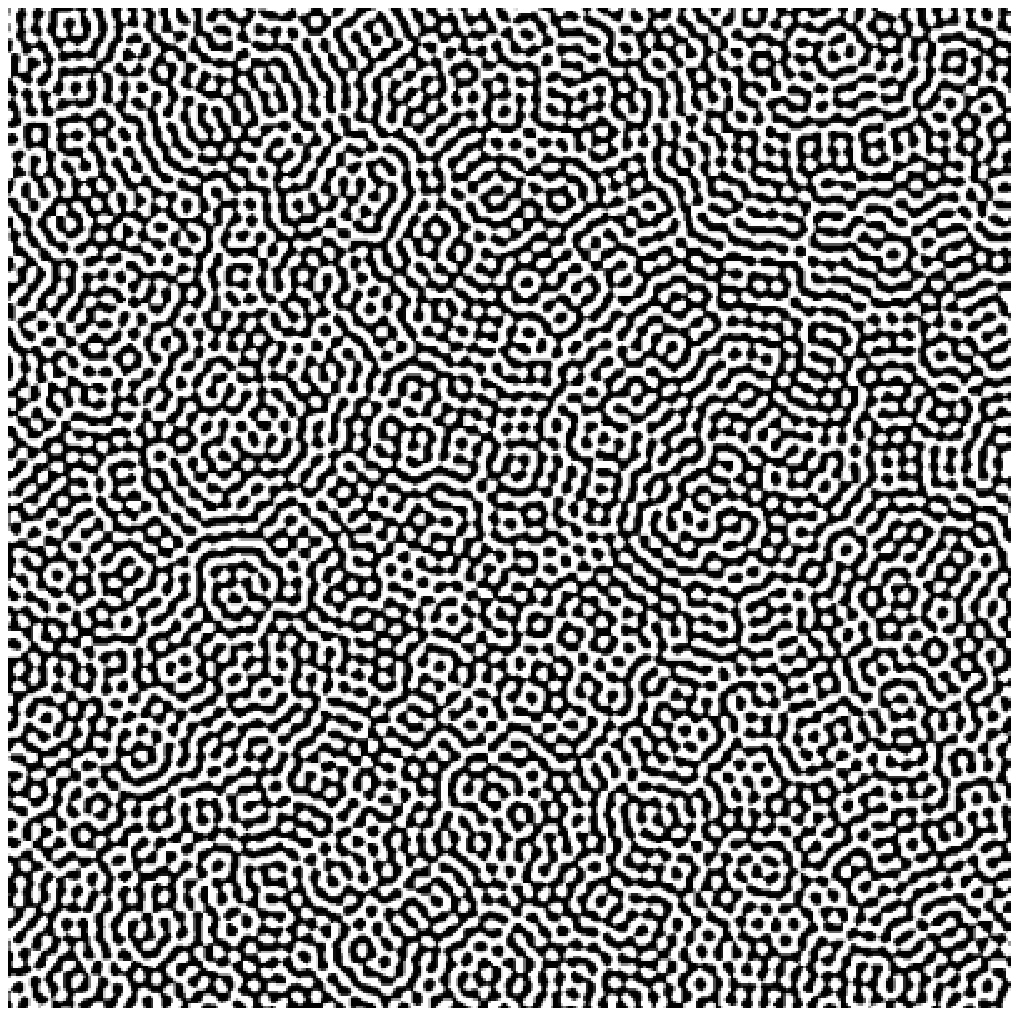} 
\caption{Left: Nodal domains of the eigenfunction of a quarter of the stadium with energy  $E=10092.029$. Right: Nodal domains of a random wave function (\ref{bessel}) with $k=100$.}
\label{fig1}      
\end{figure}     
Though  Berry's conjecture is one of the oldest conjectures in quantum chaos it has not yet been proved rigorously. Numerically it works very well. In figure~\ref{fig1} we present two pictures, the left one is a true eigenfunction of a quarter of the stadium with area equal $4\pi$  and  the right is a random realization of (\ref{bessel}) with approximately the same energy. In these figures back and white regions represent nodal domains where the function is, respectively,   positive and negative.  The two figures look very similar and the left figure was magnified in order to distinguish better the circular arc of the stadium billiard where the true wave function is zero.  

In 2002  Blum, Gnutzmann and Smilansky  attracted wide attention to such type of pictures by   initiating  the investigation of nodal domains of wave functions for different systems \cite{uzy}. 
In \cite{bogomolny}   it was argued that nodal domains of chaotic wave functions (like in figure~\ref{fig1}) can adequately be described by a critical percolation model. This conjecture gives a very detailed description of nodal domains and a large number of interesting (global) quantities can directly be transposed from percolation results (see e.g. \cite{stauffer}).  

In particular, it was shown in \cite{bogomolny} that the total number of connected nodal domains for random wave functions has  Gaussian distribution with 
mean value, $\bar{n}(E)$, and  variance, $\bar{\sigma}^2(E)$, given by the following expressions  
\begin{equation}
\frac{\bar{n}(E)}{\bar{N}(E)}=\frac{3\sqrt{3}-5}{\pi}\approx .0624
\label{mean}
\end{equation}
and 
\begin{equation}
\frac{\bar{\sigma}^2(E)}{\bar{N}(E)}=\frac{18}{\pi^2}+\frac{4\sqrt{3}}{\pi}
-\frac{25}{2\pi} \approx .0502 \;.
\label{variance}
\end{equation} 
\begin{figure}[!b]
\centering
\includegraphics[angle=-90, width=.6\linewidth]{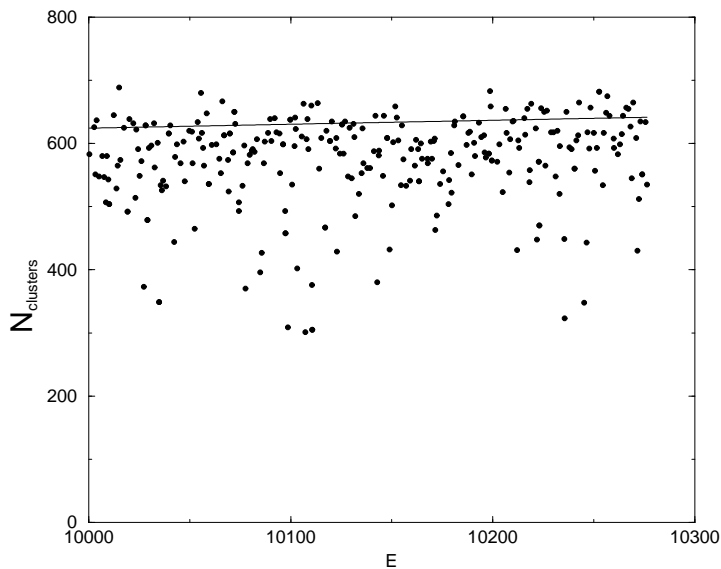}
\caption{Number of nodal clusters for stadium eigenfunctions vs energy. Solid line is  the asymptotic prediction (\ref{mean}).}
\label{fig2}
\end{figure}
Here $\bar{N}(E)$ is the mean number of eigenvalues with energy less than $E$. For billiards, $\bar{N}(E)\approx Ak^2/4\pi$ where $A$ is the billiard area.

\begin{figure}
\centering
\includegraphics[angle=-90, width=.49\textwidth]{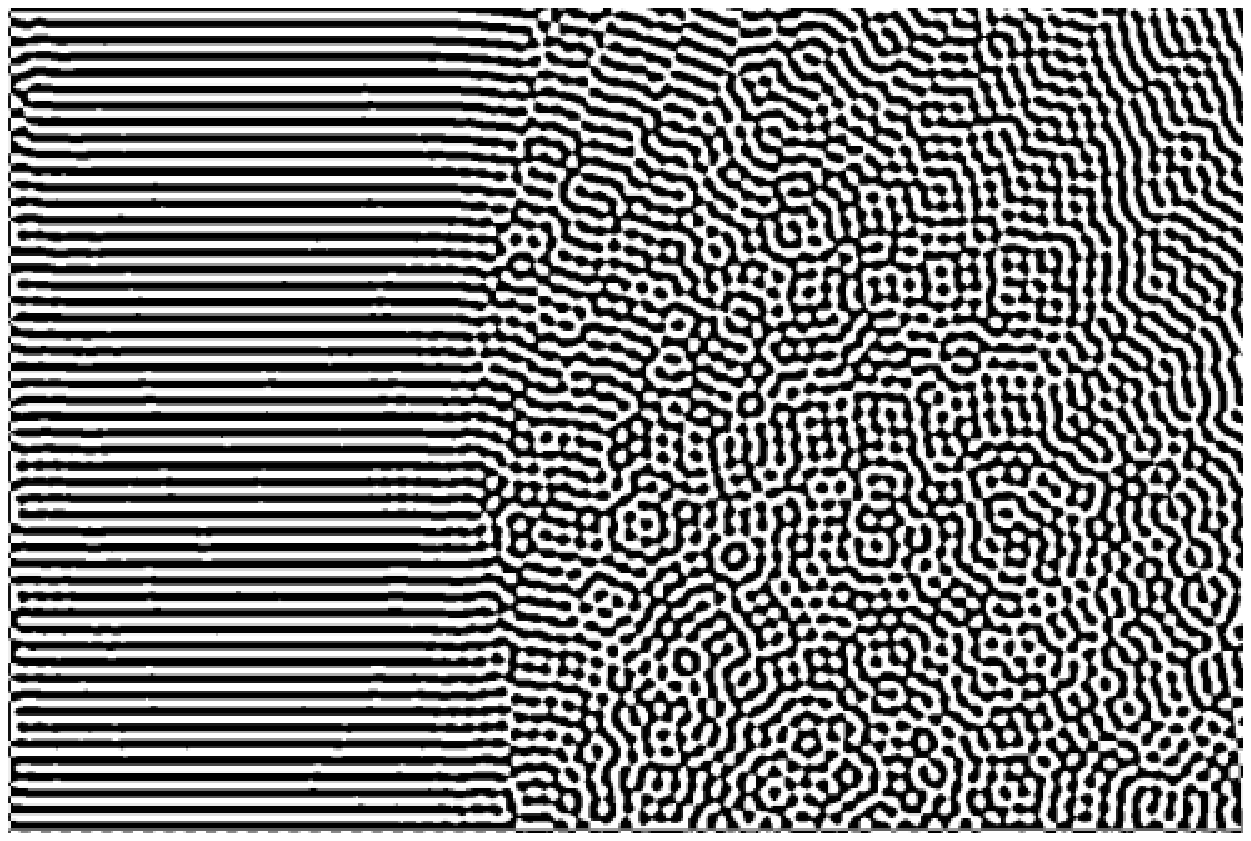}\hfill
\includegraphics[angle=-90, width=.49\textwidth]{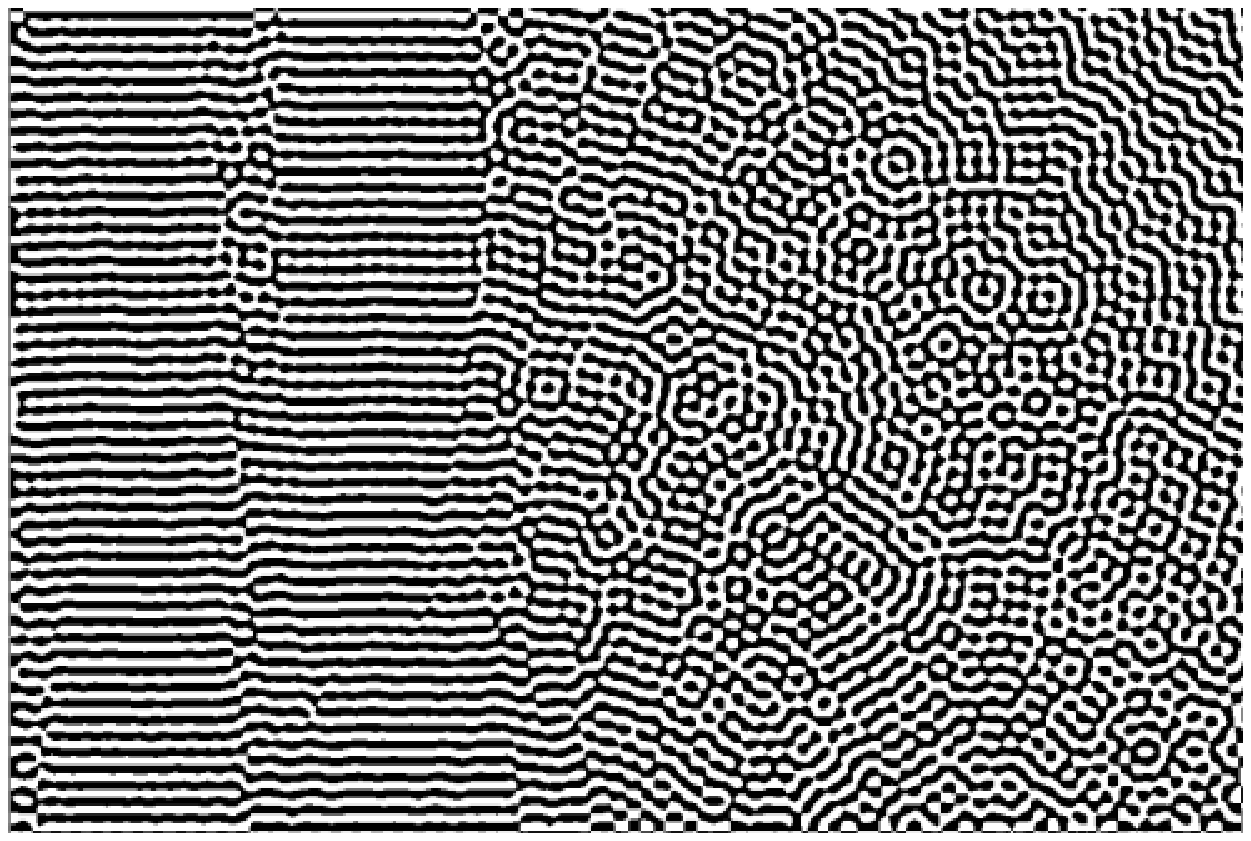}
\caption{Left:   The stadium eigenfunction with $E=10098.531$. Right: The same but  with $E=10107.147.$}
\label{fig3}
\end{figure}

	In \cite{bogomolny} it was demonstrated that these relations are well fulfilled by random wave functions. For completeness we plot in figure~\ref{fig2}  the energy dependence of  number of nodal domains computed numerically for true eigenfunctions of a quarter of the stadium close to the  $10000^{\rm th}$ level.
This figure confirms that  expression (\ref{mean}) agrees  well with the results of numerical calculations for chaotic wave functions. Still the numerical data present  deviations from the expected value (\ref{mean}) larger than it follows from (\ref{variance}). The origin of such large fluctuations is related with the existence of   wave function scars. For illustration we present  in figure~\ref{fig3}  pictures of nodal domains corresponding to the two lowest points in figure~\ref{fig2}.  It is clearly seen that these pictures have two different parts. One shows a regular pattern typical for integrable systems and the other  is practically indistinguishable from a random wave function as in figure~\ref{fig1}. It is evident that quasi-regular regions are related to the bouncing ball scar. The left eigenfunction corresponds to the single excitation of the bouncing ball  and the right one demonstrates the double perpendicular excitation.
Regions of quasi--integrable behavior have considerably less number of nodal domains than chaotic regions which explains the wide distribution of points in  figure~\ref{fig2}.      

In \cite{bogomolny} it was also checked  that the distribution of nodal domain areas agrees with  the  percolation theory prediction
\begin{equation} 
 n(s)\sim s^{-\tau}\;\;\; \mbox{ with }\tau=187/91
\label{fisher}
\end{equation}
and the  the fractal dimension of nodal clusters is close to the percolation value
\begin{equation}
D= 91/48\;. 
\label{fractal}
\end{equation}
These and other numerical calculations confirm that nodal domains of random wave functions (and of chaotic quantum systems) are correctly described by critical percolation theory. 

However, this conjecture has an intrinsic difficulty \cite{foltin}.  
The standard percolation theory deals with the following situation. One has a lattice with  each site (or edge)  positive with probability $p$ or negative with probability $1-p$. The important assumption here is that these probabilities (or concentrations) at different points are independent random variables. For random and chaotic wave functions this is not the case.  From (\ref{bessel}) it follows that
\begin{equation}
\langle \Psi(x,y)\rangle=0
\label{mean_value}
\end{equation}
and the values of the wave function at two points are correlated 
\begin{equation}
\left <\Psi(\vec{x}_1)\Psi(\vec{x}_2)\right > =\sigma^2g(|\vec{x}_1-\vec{x}_2|)
\label{thevariance}
\end{equation}
where 
\begin{equation}
g(|\vec{x}_1-\vec{x}_2|)=\sum_{m=-\infty}^{\infty} J_m(kr_1)J_m(kr_2){\rm e}^{{\rm i}m(\phi_1-\phi_2)}=J_0(k|\vec{x}_1-\vec{x}_2|)\ .
\label{two_point}
\end{equation}
In mathematical literature one often defines a Gaussian random function without explicit representation as in (\ref{berry}) but by postulating its mean value and variance (cf. (\ref{mean_value}) and (\ref{thevariance})), all other correlation functions being determined by the Wick theorem. 
 
It is well known \cite{rice} that for a Gaussian random function the probability that its value at two points has the same sign is
\begin{equation}
P(|\vec{x}_1-\vec{x}_2|)=\frac{1}{2}+\frac{1}{\pi}\arcsin G(|\vec{x}_1-\vec{x}_2|)
\end{equation}
where $G(|\vec{x}_1-\vec{x}_2|)$ is the two--point correlation function divided by its value at $0$ so that $G(0)=1$. 
Therefore, if a random Gaussian function is positive at a point, at all close-by points it will be also positive with probability close to  $1$. Only at large distances when $G(r)\to 0$ probabilities of being positive and negative become equal as in the usual percolation. 
For random wave functions $G(r)$ coincides with $g(r)$ in (\ref{two_point}) and because 
\begin{equation}
J_0(kr)\stackrel{r\to \infty}{\longrightarrow} \sqrt{\frac{2}{\pi k r}}\cos(kr-\pi/4)
\label{decay}
\end{equation}
the random wave function correlation function decays quite  slowly ($\sim r^{-1/2}$) which may destroy the validity of  percolation  theory.

One of the purpose of this paper is to show that this is not the case.  In section~\ref{harris}  we use so-called Harris' criterion \cite{harris} to demonstrate that the oscillating character of the correlation function (\ref{two_point})  leads to strong cancelations  and even the slow decay as in (\ref{decay}) is sufficient in order that  nodal domains of random wave functions be in the same universality class as non-correlated percolation. Though Harris' criterion is not mathematically rigorous,  it is widely accepted (see e.g. \cite{weinrib}) and its fulfillment is a strong argument in favor of the validity of the conjecture that nodal domains of random wave functions in the semiclassical limit $k\to\infty$ are described by critical percolation. In section~\ref{level} we generalize this conjecture to level domains i.e. regions where $\Psi(x,y)$ is bigger (or smaller) than a non-zero value $\varepsilon$. We argue that such  domains  are also  described by  percolation theory  but this time by the non-critical percolation  and the deviation from criticality is proportional to $\varepsilon$.            

\section{Harris' criterion}\label{harris}

Harris proposed his criterion in 1974 \cite{harris} to investigate the influence of random defects on the critical behavior of the Ising model. Later it was applied to many other models. In particular, in \cite{weinrib} it was used for  correlated percolation and we follow closely this paper.

Let us consider a critical percolation problem where concentrations in different sites,  denoted by $\Psi(\vec{x}\,)$,  are correlated. The criticality condition means that the mean concentration in each site equals a critical  value $p^*$
\begin{equation}
\left <\Psi(\vec{x}\,)\right >=p^* .
\end{equation}
Below  we assume that the connected part of the two--point correlation function of concentrations depends only on the difference between these points
\begin{equation}
\left < (\Psi(\vec{x}_1)-p^*)(\Psi(\vec{x}_2)-p^*)\right>=g(|\vec{x}_1-\vec{x}_2|)\ .
\label{correlation}
\end{equation}
Harris' criterion was developed to give a condition on function $g(r)$ under which correlations are unessential and critical properties of the problem under consideration are the same as for non-correlated percolation.  Harris' argumentation is as  follows \cite{harris,weinrib}.
Let $p_V$ be  the average concentration  in a finite volume $V$
\begin{equation}
p_V=\frac{1}{V}\sum_{\vec{x}\in V}\Psi(\vec{x}\,)\ .
\end{equation}
The mean value of $p_V$ over realizations equals the critical value
\begin{equation}
\left <p_V \right >=p^*
\end{equation}
but for a given realization it differs from it.

Consider the situation when inside the volume  $V$  all sites have the same concentration $p_V$. As $p_V \neq p^*$,  it corresponds to a   non-critical percolation.  From the percolation theory it is known that points are uncorrelated if they are separated by a distance larger than a certain  correlation length (the cluster size) $\xi$ 
\begin{equation}
\xi\sim |p_V-p^*|^{-\nu}
\label{xi}
\end{equation} 
where $\nu$ is certain critical index. For the standard two-dimensional percolation $\nu=4/3$ \cite{stauffer}. When $p_V$ is close to $p^*$ this correlation length is large but finite. 
Divide the whole space into cells of radius of order of $\xi$ in (\ref{xi}). Different cells can be considered as uncorrelated but  inside each cell concentrations are correlated. The natural measure of the importance of correlations is the variance, $\Delta$,  of $p_V$ which  can be computed from the knowledge of the two-point correlation function (\ref{correlation}) 
\begin{eqnarray}
\Delta &\equiv& \left <(p_V-p^*)^2\right >=\frac{1}{V^2}\sum_{\vec{x}_1,\vec{x}_2\in V}
\left <(\Psi(\vec{x}_1)-p^*)(\Psi(\vec{x}_2)-p^*)\right >\nonumber \\
&\approx& \frac{1}{V^2}\int_{\vec{x}_1\in V}\int_{\vec{x}_2\in V}
g(|\vec{x}_1-\vec{x}_2|){\rm d}\vec{x}_1{\rm d}\vec{x}_2\ . 
\label{delta}
\end{eqnarray}
Harris' criterion \cite{harris}, \cite{weinrib} states that if the variance  is small with respect to $|p_V-p^*|^2$,
\begin{equation}
\Delta\ll |p_V-p^*|^2 ,
\label{harris_crit}
\end{equation}
correlations are unessential  and all critical quantities are the same as for the standard uncorrelated  percolation. 

As the integrant in (\ref{delta}) depends only on the distance between two points one  usually simplifies  this integral  without discussion  (cf. \cite{weinrib}) by assuming that  one point  is somewhere inside $V$ and the second point is on a distance $r$ of the order of the initial cell size so that the double integral in (\ref{delta}) can be rewritten as a simple one
\begin{equation}
\Delta \approx \frac{1}{V}\int_{\vec{x}\in V}g(|\vec{x}|){\rm d}\vec{x}\ .
\label{standard_1}
\end{equation}
When the $d$-dimensional  cell $V$ is of the radius $\xi$,  Eq.~(\ref{standard_1})  leads to the following widely used expression
\begin{equation}
\Delta\approx \frac{1}{\xi^d}\int_0^{\xi}g(r)r^{d-1}{\rm d}r\;.
\label{standard_2}
\end{equation}
If  the correlation function decreases as a certain power of the distance 
 \begin{equation}
 g(r)\stackrel{r\to\infty}{\longrightarrow} r^{-\alpha}
\end{equation}
one gets
\begin{equation}
\int^{\xi}r^{-\alpha +d-1}{\rm d}r=\left \{\begin{array}{cc}
\mbox{ const} & \mbox{ when }\alpha>d\\
\xi^{d-\alpha}&\mbox{ when }\alpha<d
\end{array}
\right . .
\end{equation}
Therefore
 \begin{equation}
 \Delta\approx  \left \{\begin{array}{cc}
\xi^{-d}\sim |p_V-p^*|^{d\nu} & \mbox{ when }\alpha>d\\
\xi^{-\alpha}\sim |p_V-p^*|^{\alpha \nu} &\mbox{ when }\alpha<d
\end{array}
\right . .
\end{equation}
The comparison with (\ref{harris_crit}) leads to  the standard Harris criterion \cite{weinrib}  that   correlations are unessential provided min$(\alpha,d)\nu>2$. 

For random wave functions $g(r)$ decays as in (\ref{decay}) and because for the two--dimensional percolation  $\nu=4/3$  Harris' criterion  seems to indicate that random waves  cannot be described by usual percolation model.  But the above considerations are valid only when the correlation function $g(r)$ is positive. When it has oscillations, formulae (\ref{standard_1}) and (\ref{standard_2}) cannot be correct as by definition $\Delta$ is non-negative but these expressions may change the sign. 

The main  point is that the double integral in (\ref{delta}) cannot,  in general, be approximated by the simple integral (\ref{standard_1}).

For random wave functions correlation function is given by (\ref{two_point}) 
\begin{equation}
g(|\vec{x}_1-\vec{x}_2|)\sim  J_0(k|\vec{x}_1-\vec{x}_2|)=\sum_{m=0}^{\infty} J_{|m|}(kr_1)J_{|m|}(kr_2)
{\rm e}^{{\rm i} m (\phi_1-\phi_2)} . 
\end{equation}
Assuming for simplicity that  region $V$ is a circle,  all integrals over polar angles vanish except the one with $m=0$ and
\begin{equation}
\int_{\vec{x}_1\in V}\int_{\vec{x}_2\in V} 
g(|\vec{x}_1-\vec{x}_2|)d\vec{x}_1d\vec{x}_2=\left (2\pi \int_0^{\xi}J_0(kr)rdr\right )^2 . 
\end{equation}
Therefore  in this case
\begin{equation}
\Delta\sim \left (\frac{1}{\xi^2}\int_0^{\xi}J_0(kr)rdr\right )^2\sim \left (\frac{J_1(k\xi)}{\xi}\right )^2\sim \xi^{-3}\;.
\end{equation}
Finally as $\nu=4/3$ one concludes that 
\begin{equation}
\Delta \sim |p_V-p^*|^4\ll |p_V-p^*|^2
\end{equation}
which means that Harris' criterion is fulfilled and nodal domains of random (and chaotic) wave functions in the universal (continuum limit) have the same critical behavior as the standard critical percolation model. 

\section{Level domains and non-critical percolation}\label{level}

The success of percolation--like description of nodal domains of random wave functions naturally leads to different generalizations. In particular,  it was mentioned in \cite{bogomolny} that  level domains of random functions  can also be described by percolation theory.  But contrary to the above case one has to consider not critical but non-critical percolation theory. 

\begin{figure}
\begin{center}
\includegraphics[width=.49\linewidth, angle=270]{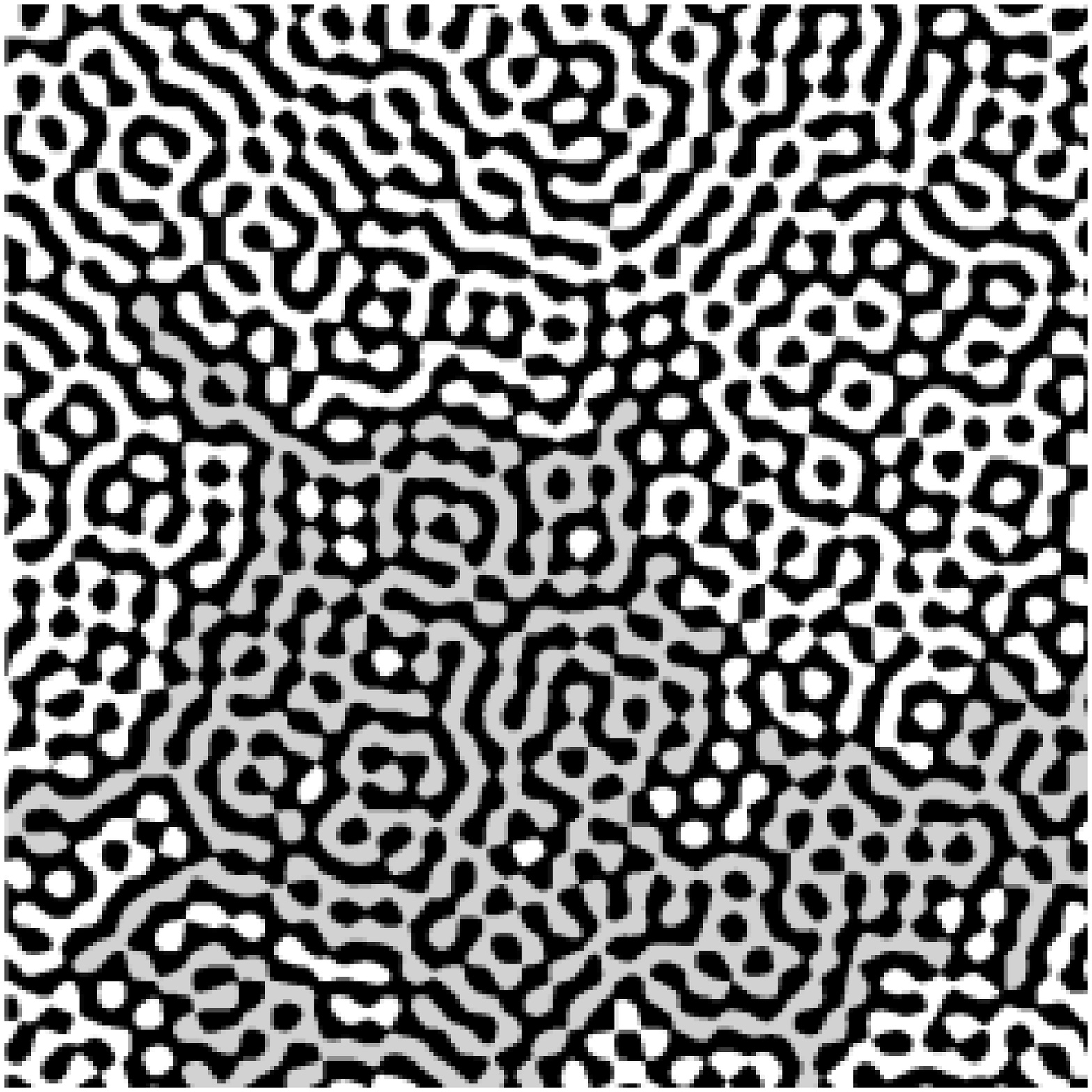}\hfill
\includegraphics[ width=.49\linewidth, angle=270]{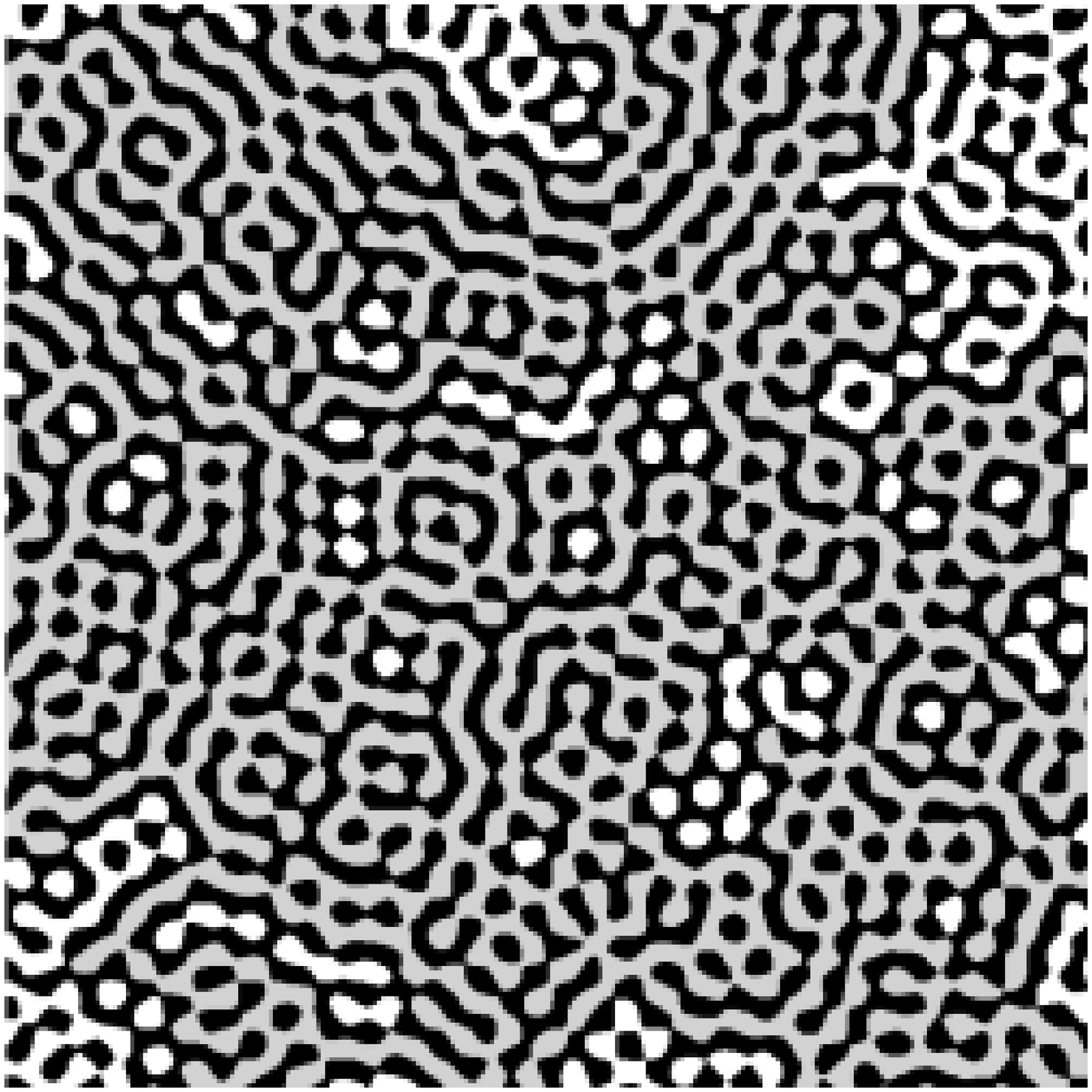}
\end{center}
\caption{Left: Nodal domains of a random wave function. Right:  Level domains of the same function with $\varepsilon=0.03$. In the both figures the largest connected clusters are highlighted.}
\label{fig4}
\end{figure}

Level domains are regions where a function is bigger than a certain value
\begin{equation}
\Psi(x,y)>\varepsilon
\label{epsilon}
\end{equation}
with $\varepsilon\neq 0$.  
To understand better how level domains look like we present in figure~\ref{fig4} two pictures. The left one displays nodal domains for one realization of a random wave function (\ref{bessel}). Boundaries between black and white nodal regions correspond to nodal lines where $\Psi(x,y)=0$. The right figure represents  level domains of the same wave function. White (resp. black) regions are domains where $\Psi(x,y)$ is bigger (resp. lower) than $0.03$ (in a certain normalization).  Boundaries between these regions (called level lines) are solutions of equation $\Psi(x,y)=0.03$. Both figures look quite similar. To stress their difference the largest connected clusters in both figures are highlighted.   In the left figure this cluster  connects different boundaries but does not cover the whole region. On the contrary, in the right figure the cluster fills practically all allowed space. This behavior is quite reminiscent of percolation transition. At critical concentration there exist an infinite cluster and a large number of  finite clusters (cf. (\ref{fisher}) and (\ref{fractal})). When the concentration is bigger than the critical,  one infinite cluster with the dimension equal to the space dimension appears.        

To quantify different relations between level domains and  non-critical percolation we use one--parameter scaling which is assumed to be valid for non-critical percolation not too far from criticality.
According to the scaling conjecture all quantity for non-critical percolation with concentration $p$ are the same as for critical percolation with concentration $p^*$ multiplied by a function depending only on one universal combination of parameters. 

For example, the number of cluster of size $s$ in critical case is given by (\ref{fisher}). For non-critical percolation it takes the following form    
\begin{equation}
n_s (p)=K_1s^{ \tau }f \left ( K_2(p-p^*)s^{\sigma}\right )
\label{scaling}
\end{equation}
where $f(z)$ is a certain  function depending on the universal argument
\begin{equation}
z=(p-p^*)s^{\sigma}\ .
\end{equation}
Here $\tau$ and $\sigma$ are universal critical indices. For two-dimensional percolation they are known analytically  (see e.g. \cite{stauffer}) 
\begin{equation}
\tau=\frac{187}{91}\;,\;\;\sigma=\frac{36}{91}\ .
\end{equation}
In (\ref{scaling})  only the constants $K_1$, $K_2$, and the threshold value $p^*$ depend on micro details of the problem such as the type of percolation (side or edge percolation) and the form of the lattice (triangular, square, etc.).  The indices $\tau$, $\sigma$ and  the function $f(z)$ are assumed to be universal (but depend on dimensionality of the  problem).

Scaling and universality are very important properties of non-critical percolation theory.  We will now check that they are also well fulfilled for level domains of random wave functions.
First we need a relation between the deviation from criticality, $p-p^*$,  and $\varepsilon$ in (\ref{epsilon}).  To simplify the formulas below it is convenient to normalize billiard wave functions by the  condition
\begin{equation}
 \int_A|\Psi(x,y)|^2dxdy=A
 \end{equation}
 where $A$ is the billiard area.
 
With  this normalization  random wave functions (\ref{bessel}) are Gaussian random functions with $\sigma=1$ (cf. (\ref{two_point})). Therefore, the probability density that $\Psi(x,y)=t$ is 
 \begin{equation}
P(t)=\frac{1}{\sqrt{2\pi}}{\rm e}^{-t^2/2}
\end{equation}
and the probability $P_{>}(\varepsilon) $ that $\Psi(x,y)>\varepsilon$ takes the form
\begin{equation}
P_{>}(\varepsilon)=\int_{\varepsilon}^{\infty}P(t){\rm d}t\approx \frac{1}{2}-
\frac{\varepsilon}{\sqrt{2\pi}} +{\cal O}(\varepsilon^3)\ .
\end{equation}
Consequently, for small $\varepsilon$  
\begin{equation}
p-p_c\sim \varepsilon
\end{equation}
and for level domains the scaling relations as in (\ref{scaling}) are valid with the substitution $p-p^*\longrightarrow \varepsilon$.

Consider first the total number of connected level domains.
For this quantity there exists no explicit scale as in (\ref{scaling}) and the scaling prediction is particular simple. For the non-critical percolation
\begin{equation}
\frac{\bar{n}(p)}{\bar{n}(p^*)}=f(p-p^*)
\label{n_perc}
\end{equation}
so for level domains it should be
\begin{equation}
\frac{\bar{n}(\varepsilon)}{\bar{n}(0)}=f(K\varepsilon)
\label{n_level}
\end{equation}
with a certain constant $K$ and the same function $f(z)$ . Here $\bar{n}(p)$ is the mean number of connected clusters for a percolation model with concentration $p$. $\bar{n}(\varepsilon)$ is the same quantity but for level domains (\ref{epsilon}). $\bar{n}(0)$ is the mean number of nodal domains given by (\ref{mean}). 

\begin{figure}
\begin{center}
\includegraphics[ angle=270,  width=.7\textwidth]{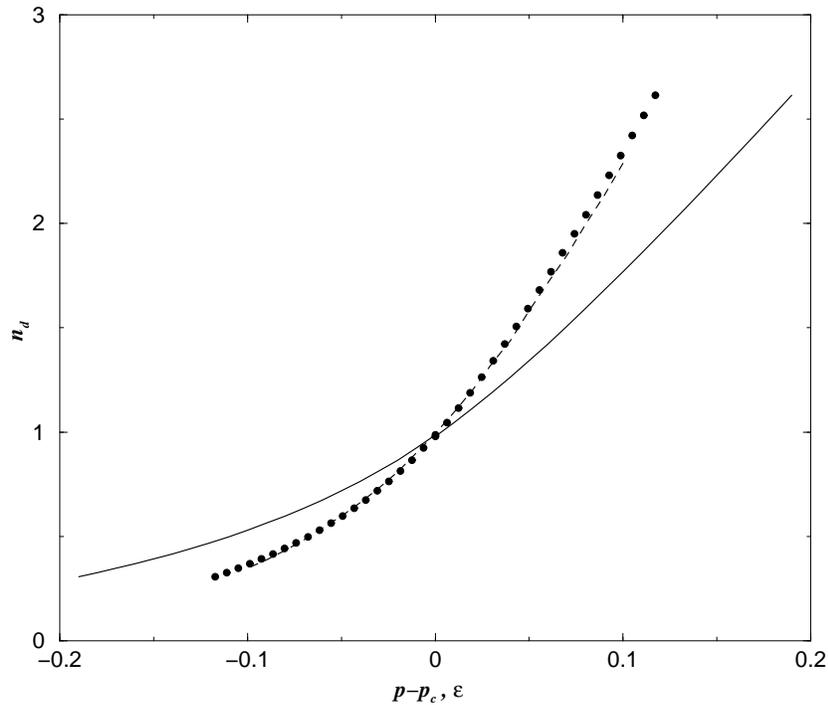}
\end{center}
\caption{Ratio of the number of clusters to its critical value for level domains of random wave functions (solid line) and for non-critical bond percolation on a square lattice. For the first curve the abscissa axis is $\varepsilon$, for the second one it represents $p-p^*$.  Black circles are results of rescaling $p-p^*=\varepsilon/1.62$.}
\label{fig5}
\end{figure}

In figure~\ref{fig5} there are two curves. The solid line is the ratio (\ref{n_level}) computed numerically for random Gaussian functions. Data are  well fitted by a quadratic fit  $1+.6135\varepsilon+13.47\varepsilon^2$. The dashed line represents the ratio (\ref{n_perc}) computed numerically for bond percolation on $110\times 110$ square lattice. Black circles  result from the  rescaling $p-p^*=\varepsilon/1.62$. It is clearly seen that points for level domains after rescaling agree very well with data from pure percolation which confirms the scaling (\ref{n_perc}) and (\ref{n_level}).  

\begin{figure}
\begin{center}
\includegraphics[ angle=270, width=.7\textwidth]{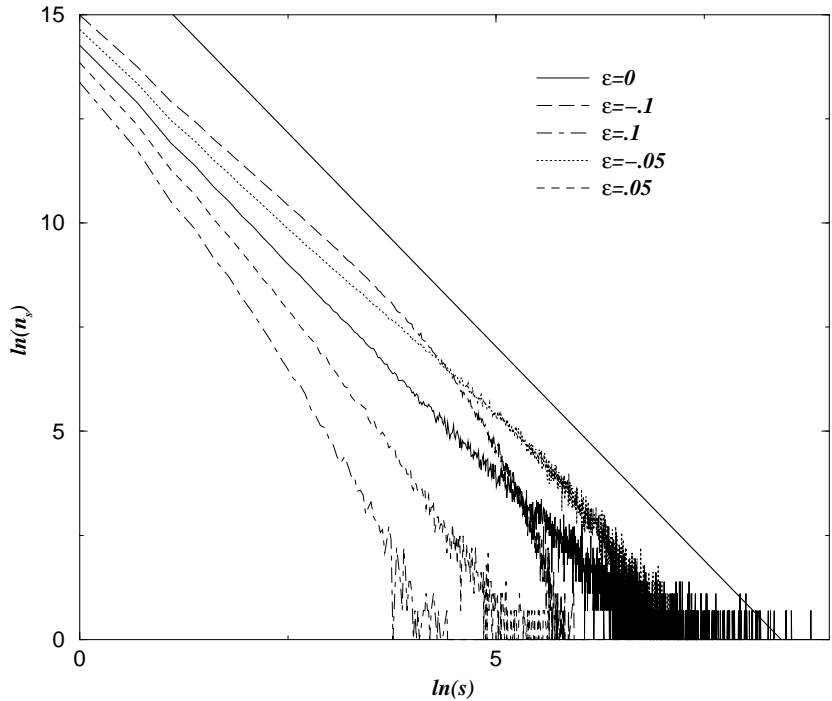}
\end{center}
\caption{Number of domains of area $s$ for random wave functions. Different curves represent results for different values of $\varepsilon$. Straight line has the slope $ -187/91$ predicted for nodal domains.}
\label{fig6}
\end{figure}

\begin{figure}
\begin{center}
\includegraphics[angle=270,  width=.495\textwidth]{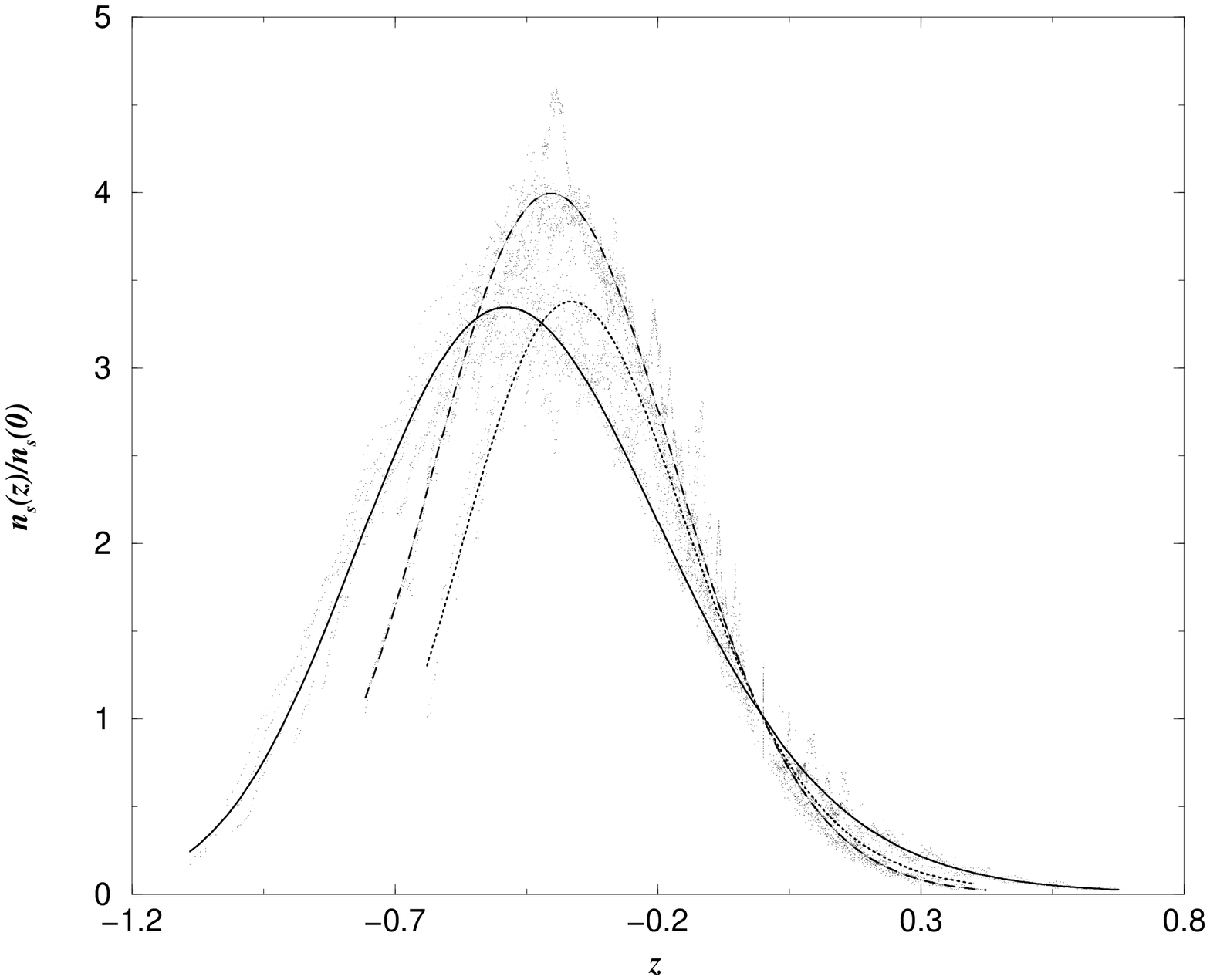}\hfill
\includegraphics[ angle=270,  width=.495\textwidth]{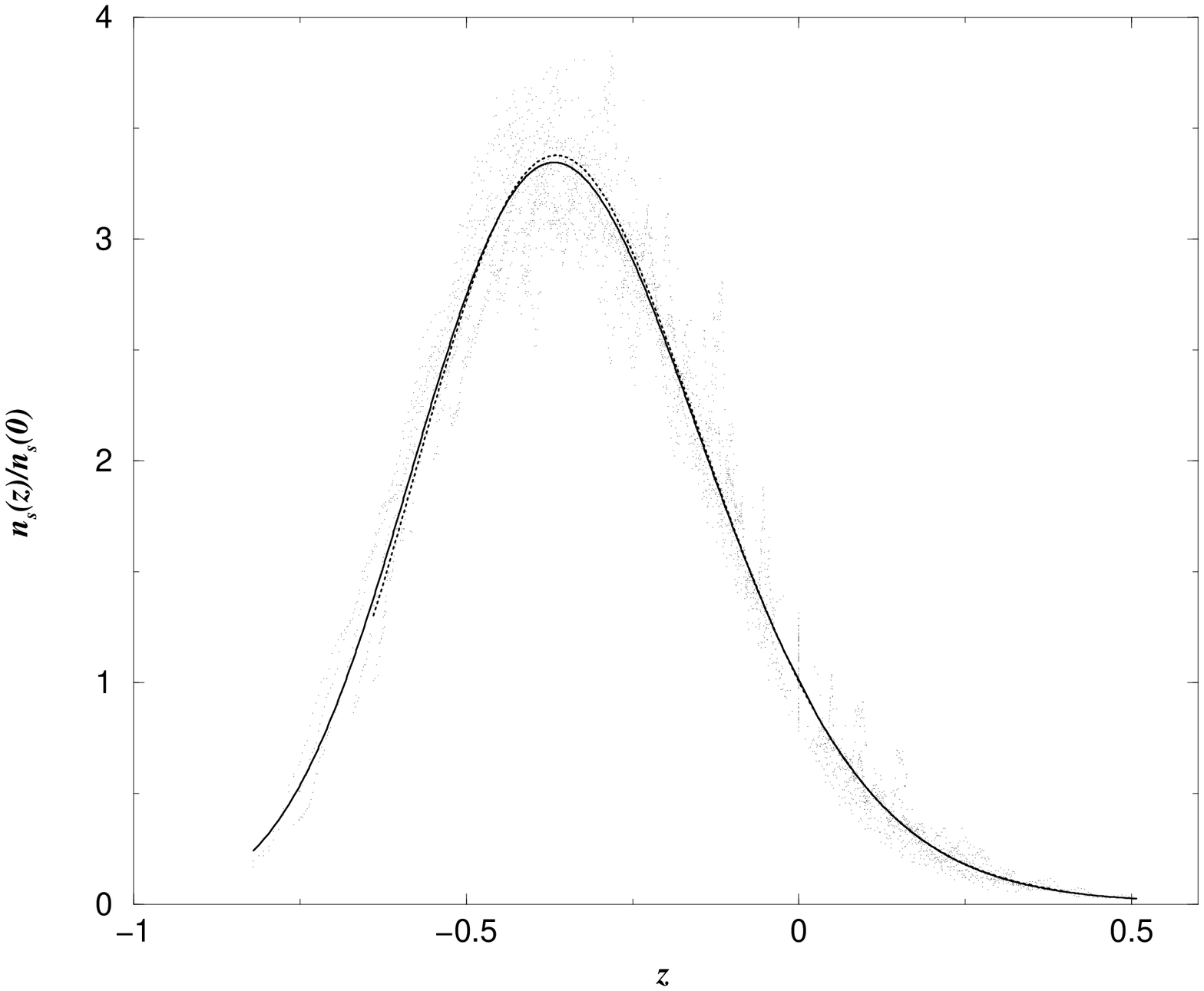}
\end{center}
\caption{Left: Scaling relations. Solid line -- random wave functions with $k=200$.
Dashed line -- bond percolation on a square $110\times 110$ lattice.
 Dotted line -- the same but on a $50\times 50$ lattice. Right:  Rescaling of dotted curve by $p-p^*=\varepsilon/1.33$.}
\label{fig7}
\end{figure}

\begin{figure}
\begin{center}
\includegraphics[ angle=270,  width=.7\textwidth]{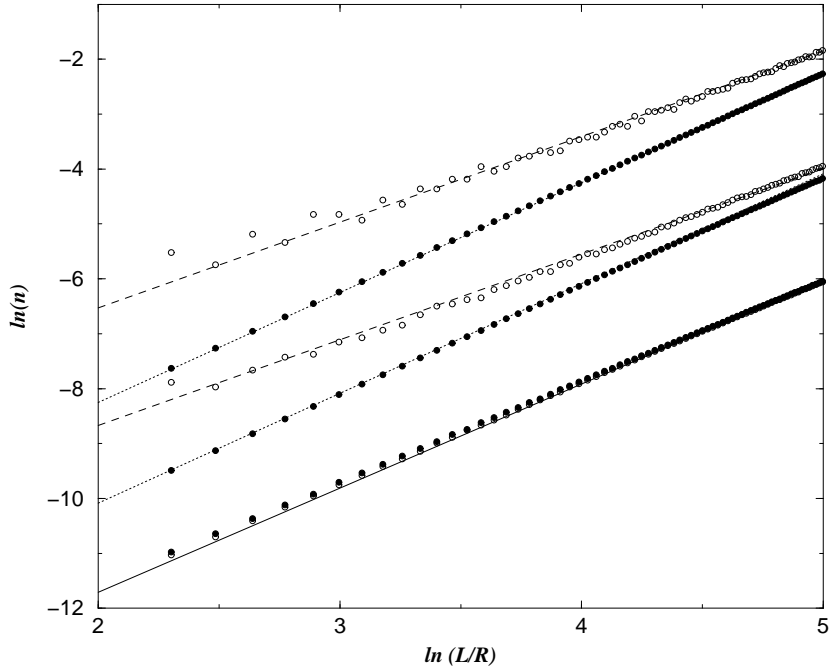}
\end{center}
\caption{Fractal dimension of level domains. From bottom to top: $\varepsilon=0$, $\varepsilon=0.1$, and $\varepsilon=0.2$. For all curves $k=200$. The slope of the solid line is $91/48$, of the dotted line is $2$, of the dashed line it is $1.56$. Black circles represent  domains where  $\Psi(x,y)<\varepsilon$ and open ones regions with $\Psi(x,y)>\varepsilon$. }
\label{fig8}
\end{figure}

In figures~\ref{fig6} -- \ref{fig7} we present the verification of (\ref{scaling}) for level domains of random wave functions.  In figure~\ref{fig6} the number of clusters with area $s$ is plotted for a few different values of $\varepsilon$ and $k=200$.  Curves are clearly different and slopes for $\varepsilon\neq 0$ differ from  the critical percolation value (\ref{fisher}) predicted for nodal domains.  Then we rescale these data using as the abscissa axis the variable $z=\varepsilon s^{\sigma}$ as it follows from the one-parameter scaling. The resulting points are concentrated close to each other and  in the left of figure~\ref{fig7} the solid line indicates the average  of all such calculations which supports the existence of scaling for level domains. In the same figure two more curves are presented. The dashed line represents the results of similar calculations but for the non-critical bond percolation on a square $110\times 110$ lattice and the dotted line is the same but on a square lattice with $50\times 50$ sites.  For these curves the  $z$ variable is defined as in (\ref{scaling}), namely, $z=(p-p^*)s^{\sigma}$. Once more the scaling relation is well satisfied.  Finally on the right part of  figure~\ref{fig8} it is shown  that  after the rescaling  $p-p^*=\varepsilon/1.33$, the curve of level domains practically coincides with the one for non-critical percolation on a square $50\times$ lattice. 
 
\begin{center}
{\bf \ \large }\end{center}

Finally in figure~\ref{fig8} the results of numerical calculations for fractal dimensions of clusters are presented. According to  percolation theory (see e.g. \cite{stauffer}) the dominant cluster  has the dimension  $2$ but the other ones have the fractal  dimension equal approximately to $1.56$. Notice that for critical percolation (and the nodal domains) the fractal dimension of clusters is $91/48$.

\section{Conclusion}

The conjecture that random (and chaotic) wave functions in the semiclassical limit can be described by percolation theory is analyzed.  We demonstrate that, though wave correlations decay slowly, a  careful application of Harris' criterion suggests that nodal domains of random wave functions are indeed in the same universality class as the critical non-correlated percolation.  We also present arguments in favor that  the non-critical percolation is useful in the investigation of non-zero level domains of random wave functions.  Let us mention that in \cite{remy} it is shown that nodal lines of random wave functions are  close to SLE$_6$ curves which describe  boundaries of critical percolation clusters. All these results confirm the above conjecture. It is of interest to find a rigorous proof of it.       

\section*{Acknowledements}

E.B.  is  grateful to G. Foltin, U. Smilansky, and J. Keating for stimulating  discussions and to O. Bohigas for a critical reading of the manuscript.

\end{document}